\documentclass[aps,showpacs,amsmath,eqsecnum,nofootinbib,12pt]{revtex4}

\usepackage{amsfonts}
\usepackage{amssymb}
\usepackage{color}
\usepackage{mathrsfs}

\def\bc{\begin{center}}
\def\nno{\nonumber}
\def\ec{\end{center}}
\def\be{\begin{eqnarray}}
\def\ee{\end{eqnarray}}

\newcommand{\omits}[1]{}

\definecolor{dyellow}{rgb}{1.,0.8,.0}
\definecolor{myblue}{rgb}{.1,.1,.7}
\definecolor{dcyan}{rgb}{.0,.6,.6}
\definecolor{dmagenta}{rgb}{0.6,0.0,0.6}
\definecolor{brown}{rgb}{0.6,0.2,0.}
\definecolor{darkblue}{rgb}{.0,.0,0.5}
\definecolor{darkred}{rgb}{0.75,0.0,0.0}
\definecolor{orange}{rgb}{1.,.6,.0}
\definecolor{dorange}{rgb}{0.8,.4,.0}
\definecolor{lightgray}{rgb}{0.7,0.7,0.7}
\definecolor{darkgreen}{rgb}{0.0,0.6,0.0}
\definecolor{purple}{rgb}{.4,.0,.4}

\def\Om{\Omega}

\def\eps{\epsilon}

\def\om{\omega}

\def\d#1#2{\frac{\displaystyle #1}{\displaystyle #2}}
\def\r{\partial}

\newcommand{\dS}{$d{S}$}
\newcommand{\Mink}{${M}ink$}
\newcommand{\AdS}{${A}dS$}
\newcommand{\BdS}{${B}d{S}$}

\newcommand{\SR}{special relativity}

\newcommand\btd{\raise 2pt
\hbox{$\hat\bigtriangledown$}\hskip 1.5pt}
\newcommand\bt{\raise 2pt
\hbox{$\bigtriangledown$}\hskip 1.5pt}

\def\cA{{\cal A}}
\def\cB{{\cal B}}

\def\PRD{{\it Phys. Rev.}~{\bf D}}

\def\PLA{{\it Phys. Lett.}~{\bf A}}

\def\CTP{{\it Comm. Theor. Phys. }}

\def\no{\noindent}

\begin{document}

\title{Yang's Model as Triply Special Relativity and\\
the Snyder's Model--de Sitter Special Relativity Duality}

\author{{Han-Ying Guo}$^{1}$}
\email{hyguo@itp.ac.cn}
\author{{Chao-Guang Huang}$^{2}$}
\email{huangcg@mail.ihep.ac.cn}
\author{{Hong-Tu Wu}$^{3}$}
\email{lobby_wu@yahoo.com.cn}

\affiliation{%
${}^1$ Institute of Theoretical Physics, Chinese Academy of
Sciences, Beijing 100080, China,}
\affiliation{%
${}^2$  Institute of High Energy Physics, Chinese Academy of
Sciences, P.O. Box 918-4, Beijing 100049, China,}
\affiliation{%
${}^3$Department of Mathematics, Capital Normal University, Beijing
10037, China.}

\date{March, 2008}

\begin{abstract}
We show that if Yang's quantized space-time model is completed at
both classical and quantum level, it should contain both Snyder's
model, the de Sitter \SR\ and their duality.
\end{abstract}


\maketitle

\tableofcontents

\section{Introduction}
Some sixty years ago, Snyder \cite{Snyder} proposed a quantized
space-time model by means of the projective geometry approach to the
de Sitter (\dS)-space of momentum
 with two
universal constants: $c$ and $a$, a scale near {or at} the Planck
length $\ell_P$. The  4-energy-momentum were
 defined by the inhomogeneous projective coordinates. Then,
Snyder identified  the space-time coordinates'  noncommutative
operators $\hat x^\mu$ with 4-`translation' generators of a
\dS-algebra $\mathfrak {so}(1,4)$ and other operators as angular
momentum for an $\mathfrak {so}(1,3)\subset \mathfrak {so}(1,4)$.

Soon after, Yang \cite{Yang} extended Snyder's model to the one with
the third constant, the radius $R$ of a \dS\ universe in order to
recover the translation under $R\to \infty$. Yang found an
$\mathfrak {so}(1,5)$ algebra with $c, a$ and $R$ in a 6-d space
with Minkowski (\Mink) signature. In Yang's algebra, there are two
$\mathfrak {so}(1,4)$ subalgebras for coordinate operators $\hat
x^\mu$ and momentum operators $\hat p^\mu$, respectively,  with a
common $\mathfrak {so}(1,3)$ for angular momentum operators $\hat
l^{\mu\nu}$. And  the algebra is invariant under a $Z_2$ dual
transformation between $a, \hat x^\mu$ and $\hbar /R, \hat p^\mu$.
This is a UV-IR parameters' transformation.

Recently, the `doubly special relativity' or the `deformed special
relativity' (DSR) has been proposed \cite{DSR}. There is also a
universal constant $\kappa$ near the Planck energy, related to
$\hbar/a$ in Snyder's model in addition to $c$.  In the sense with
one more universal constant $a$ near {or at} the Planck length
$\ell_P$ in addition to $c$, Snyder's model might be regarded as the
earliest DSR and some of DSR models can be given as the
generalization of Snyder's model \cite{DSRdS}. Soon,
 the `triply special
relativity' (TSR) \cite{TSR} has also been proposed under a deformed symmetry
with one more universal constant, the universe radius $R \sim \Lambda^{-1/2}$.
Later, it is found that the Lie algebra form of the deformed symmetry
 is just Yang's algebra in the 6-d space \cite{3SR}. Thus, Yang's
 model might  be regarded as a TSR (see also \cite{Kong, QS}).

The projective geometry approach to \dS-space is basically
equivalent to the Beltrami model  of \dS-space (\BdS). It is
important to emphasize that the Beltrami coordinates or
inhomogeneous projective ones, without the antipodal identification
for preserving orientation, play an important role in analogy with
the \Mink-coordinates in a \Mink-space. Namely, in the Beltrami
atlas, particles and light signals move along the time-like or null
geodesics {being straight world-lines} with {\it constant}
coordinate velocities, respectively. Among these systems, the
properties are invariant under the fractionally linear
transformations with a common denominator of \dS-group. If these
motions and  systems could be regarded as of inertia without
gravity, there should be the
 principle of inertia in \dS/\AdS-spacetime, respectively.

In fact, just as weakening the fifth axiom leads to non-Euclidean
geometry, giving up Einstein's Euclidean assumption on the rest
rigid ruler and clock in \SR\ leads to two other kinds of special
relativity on the \dS/\AdS-spacetime with radius $R$. They are based
on the principle of inertia and the postulate of universal constants
($c, R$) on an almost equal footing with
 the \SR\ on \Mink-spacetime of $R \to \infty$ 
\cite{BdS, IWR, BdS05, T, NH, Lu05, PoI, Lu80, Yan}.

On the other hand, it is interesting to see \cite{PRdual, Duality}
that in terms of the Beltrami model of \dS-space \cite{BdS,BdS05},
there is an important one-to-one correspondence between Snyder's
quantized space-time model \cite{Snyder} as a DSR \cite{DSR,DSRdS}
and the \dS\, special relativity \cite{IWR, BdS, BdS05, T}.
Actually, the \dS\ special relativity can be regarded and simply
formulated as a spacetime-counterpart of Snyder's model for
\dS-space of momentum so long as the constant $\hbar/a$ in Snyder's
model as \dS-radius of momentum near the Planck scale is replaced by
$R$ as radius of \dS-spacetime. Inspired by the correspondence, the
Snyder's model--\dS\ \SR\ duality as a UV-IR duality is proposed
\cite{PRdual, Duality}.

Since Snyder's quantized 4-d space-time model is on a 4-d \dS-space
of momentum, if Yang's model is really generalized Snyder's with the
third universal constant $R$, it should also be back to 4
dimensions. But, how to realize Yang's model in 4 dimensions
completely?  In his very short paper, Yang did not answer the
question.  Recall that there are three 4-d maximally symmetric
spacetimes with maximum symmetries of ten generators, which are just
the \Mink/\dS/\AdS-space with $ISO(1,3)/SO(1,4)/SO(2,3)$ invariance,
respectively. Thus, it is impossible to realize Yang's
$\mathfrak{so}(1,5)$ algebra with fifteen generators on one space of
4-dimensions in the sense of  Riemann geometry and Lie symmetry. The
TSR realization of Yang's $\mathfrak{so}(1,5)$ algebra gives a
tentative 4-d realization. But, it is in terms of a deformed algebra
with  non-commutative geometry.  The fact that there are two
$\mathfrak{so}(1,4)$ subalgebras with a common homogeneous Lorentz
algebra $\mathfrak{so}(1,3)$ in Yang's $\mathfrak{so}(1,5)$ suggests
another kind of realization: A pair of \dS-spaces of 4-dimensions
with a dual relation.

In this paper we show that if Yang's model can be completed with
such a kind of 4-d realizations at both classical and quantum level,
this complete Yang model should contain both Snyder's quantized
space-time model, the \dS\ \SR\ and their duality.

This paper is arranged as follows. In section \ref{yang}, we first
recall and complete Yang's model with a UV-IR dual invariance in a
6-d dimensionless \Mink-space at both classical and quantum level.
Then, in section \ref{snyderdssr}, we show that the two
$\mathfrak{so}(1,4)$ subalgebras in the complete Yang model relevant
to the space-time coordinate operators $\hat x^\mu$ and the momentum
operators $\hat p^\mu$ are the same as Snyder's $\mathfrak{so}(1,4)$
algebra of quantized space-time and the algebra for `quantized'
energy, momentum, and angular momentum in a \dS-space of spacetime,
respectively. We also present a way to get Snyder's model, the \dS\
\SR\ and their duality from the Yang model.  Finally, we end with
some concluding remarks.

\section{A Complete Yang Model and a UV-IR Duality}\label{yang}

Under Yang's $\mathfrak{so}(1,5)$ algebra, there is an invariant 
quadratic form of signature $-4$ \cite{Yang} in a 6-d dimensionless
\Mink-space $\mathscr{M}^{1,5}$.
 Then, the metric in $\mathscr{M}^{1,5}$ reads
\begin{equation}\label{dChi}%
d\chi^2=\eta_{\cA \cB}^{} d\zeta^{\cA} d\zeta^{\cB},\quad
\cA,\cB=0,\ldots,5,
\end{equation}
where $\eta_{\cA \cB }^{}={\rm diag}(+, -, -, -, -, -)$. The dimensionless
 canonical `momentum' conjugate to the dimensionless `coordinate' $\zeta^\cA$
can be introduced as $N_{\cA}=\eta_{\cA\cB}^{}\frac{d
\zeta^{\cB}}{d\chi}.$ Thus, there is a 12-d phase space $({\cal M},
\Omega)$ with a symplectic structure $\Om$ and the non-vanishing
basic
 Poisson bracket in  ($\zeta^{\cA},N_{\cA}$):
$\{\zeta^{\cal{A}},N_{\cal{B}}\}=-\delta^{\cal{A}}_{\cal{B}}.$
Obviously, the dimensionless 6-`angular momentum' %
 ${\mathscr L}^{\cal AB}:=\zeta^{\cal A} N^{\cal B}-\zeta^B N^{\cal A}
 $ as the classical counterpart of Yang's operators (see below)
 form an $\mathfrak{so}(1,5)$ algebra under Poisson bracket:
 \begin{equation}\label{so15}
 \{{\mathscr L}^{\cal AB}, {\mathscr L}^{\cal CD}\} =\eta^{\cal
AD}{\mathscr L}^{\cal BC}+\eta^{\cal BC}{\mathscr L}^{\cal
AD}-\eta^{\cal AC}{\mathscr L}^{\cal BD}-\eta^{ \cal BD} {\mathscr
L}^{\cal AC}.
\end{equation}

Under canonical quantization, in the `coordinate' picture with
$\hat{N}_{\cal{A}}=i  \frac{\partial }{\partial
\zeta^{\cal{A}}},~[\hat{\zeta}^{\cal{A}}, \hat{N}_{\cal{B}}]=-i
 \delta^{\cal{A}}_{\cal{B}}$,  they become operators
$\hat{\mathscr L}^{\cal AB}$ forming the algebra under
Lie bracket.
Now,  the following operators  are  just the operators in Yang's
model \cite{Yang} up to some redefined
coefficients
\begin{equation}
\begin{split}
\hat{x}_{0} & =i a\left(\zeta^5 \frac{\partial }{\partial \zeta^0}+
\zeta^0 \frac{\partial }{\partial \zeta^5} \right)=a  \hat{\mathscr{L}}^{50} , \\
\hat{x}_{i} & =i a \left(\zeta^5 \frac{\partial }{\partial \zeta^i}-
\zeta^i \frac{\partial }{\partial \zeta^5} \right)=- a \hat{\mathscr{L}}^{5i} , \\
\hat{p}_{0} & = \frac{i\hbar}{R}\left(\zeta^4 \frac{\partial}{\partial \zeta^0}
+\zeta^0 \frac{\partial }{\partial \zeta^4}\right)
=\frac {\hbar} R \hat{\mathscr{L}}^{40} , \\
\hat{p}_{i} &= \frac{i\hbar}{R}\left(\zeta^4 \frac{\partial
}{\partial \zeta^i}-\zeta^i \frac{\partial }{\partial
\zeta^4}\right)=-\frac {\hbar} R \hat{\mathscr{L}}^{4i} ,  \\
\hat{M}_{i}&=i \hbar\left(\zeta^0\frac{\partial }{\partial \zeta^i}+\zeta^i \frac{\partial}{\partial \zeta^0}\right)
= - {\hbar}\hat{\mathscr{L}}^{0i}, \\
\hat{L}_{i}& = i\hbar {\eps_{i}^{\ jk}\left(\zeta_{j} \frac{\partial
}{\partial \zeta^{k}}\right)} =\frac {\hbar} 2
\eps_{ijk}\hat{\mathscr{L}}^{jk},\\%
 \hat \psi &= i\frac a R \left
(\zeta^5\d{\r}{\r\zeta^4}-\zeta^4\d{\r}{\r\zeta^5}\right
) =\frac a R \hat{\mathscr{L}}^{45}
\end{split}
\label{yang operators}
\end{equation}
with $\eps_{123}=\eps_{1}^{\ 23}=1$ and
$\zeta_j=\eta_{j\cal A}^{}\zeta^{\cal A}$.
They form Yang's $\mathfrak{so}(1,5)$ algebra as follows:
\begin{equation}\label{yang so15}%
\begin{split}
[\hat p^\mu, \hat p^\nu]=i\hbar R^{-2} \hat l^{\mu\nu},&\quad
[\hat l^{\mu\nu}, \hat p^\rho]=i\hbar(\eta^{\nu \rho}\hat
p^\mu-\eta^{\mu\rho}
\hat p^\nu), \quad \hat p ^\mu =\eta^{\mu\nu}\hat p_\mu,
\quad \hat l^{\mu\nu}=\hbar \hat {\mathscr{L}}^{\mu\nu} \nno \\
[\hat x^\mu, \hat x^\nu]=i\hbar^{-1} a^2 \hat l^{\mu\nu},&\quad [\hat
l^{\mu\nu}, \hat x^\rho]=i\hbar(\eta^{\nu \rho} \hat x^\mu-\eta^{\mu
\rho}
\hat x^\nu),\quad  \hat x ^\mu =\eta^{\mu\nu}\hat x_\mu\\\no
[\hat x^\mu,\hat p^\nu]=i{\hbar} \eta^{\mu\nu}\hat \psi,&\quad
[\hat \psi, \hat x^\mu ]=-i a^2 {\hbar ^{-1}} \hat p^\mu,\\
\no[\hat\psi,\hat p^\mu ]= i\hbar R^{-2}\hat x^\mu ,&\quad  [\hat
\psi, \hat l^{\mu \nu}] = 0, \nno
\end{split}
\end{equation}
together with an $\mathfrak{so}(1,3)$ for the 4-d angular momentum
operators.

It is clear that there are two $\mathfrak{so}(1,4)$ for coordinate
operators $\hat x^\mu$ and momentum operators  $\hat p^\nu$,
respectively, with a common $\mathfrak{so}(1,3)$ for $ \hat
l^{\mu\nu}$. It is also clear that in Yang's algebra with respect to
the 6-d `angular momentum' there is  a $Z_2=\{e, r|r^2=e\}$ dual
transformation with
\be \label{duality}\no%
r: \qquad a \rightarrow \d {\hbar} R, \quad \hat x^\mu
\rightarrow \hat p^\mu, \quad  \hat \psi \rightarrow -\hat\psi.
\ee %
Since $a$ is near or equal to the Planck length $\ell_P$ and $R$
is the radius of a \dS\ universe, the invariance under the $Z_2$
dual transformation is a UV-IR duality.

\section{The Snyder's Model--\dS\ \SR\ Duality from the Yang Model}\label{snyderdssr}

\subsection{Snyder's model from the Yang model}

Snyder considered a homogeneous quadratic form
$-\eta^2=\eta^2_0-\eta^2_1-\eta^2_2-\eta^2_3-\eta^2_4
:=\eta^{AB}\eta_A^{}\eta_B^{} <0,$ partially inspired by Pauli.  It is a model
via homogeneous (projective) coordinates of a 4-d momentum space of constant
curvature, a \dS-space of momentum. In fact, it can also be started from  a
\dS-hyperboloid ${\cal H}_a$ in a 5-d \Mink-space of momentum with radius
$1/a$
\be\label{Ha} %
&{\cal H}_a:\quad { \eta^{AB} \eta_A^{} \eta_B^{}
=-\d{ \hbar^2}{a^2},
\quad ds_a^2 =\eta^{AB}d\eta_A^{} d\eta_B^{}.} \\
\nno  %
\ee

Snyder's inhomogeneous projective momentum is almost the same as the
 momentum in Beltrami coordinates. In order to preserve orientation, the antipodal
identification should not be taken so that the Beltrami atlas should
contain at least eight patches to cover the hyperboloid (see, e.g.
\cite{BdS}). In the patch $U_{4+}, { \eta_4}>0 $, Snyder's Beltrami
momentum read
\begin{equation}
{q_\mu^{}=\frac{\hbar}{a}  \frac{\eta_\mu^{}}{\eta_4^{}}}.
\end{equation}
Now the metric in the patch becomes $ds_a^2={g_{a}^{\mu \nu} dq_\mu^{} dq_\nu^{}}$ with%
 \be\label{dsa}%
{ g_{a}^{\mu \nu}=\sigma^{-1}(q)\eta^{\mu \nu}+\d {a^2} { \hbar
^2}q^\mu q^\nu \sigma^{-2}(q)}, \quad \sigma(q)=1-\d
{a^2}{\hbar^2} q^\nu q_\nu>0,
 \ee%
where ${q^\mu =\eta^{\mu\nu} q_\nu^{}}.$ Along geodesic that is the
great `circle' on ${\cal H}_a$, the spacetime `coordinates' and
angular momentum 
are conserved%
\begin{align}\label{xaC}
x_a^\mu&={ R}\sigma^{-1}(q) \frac{d q^\mu}{ds_a}=consts,\\\nno
l_a^{\mu \nu}&={ R}(q^\mu \frac{dq^\nu}{ds_a}-q^\nu
\frac{dq^\mu}{ds_a})=consts.
\end{align}
Importantly, from these conserved Killing observables and
${q_0^{}}=E$ as energy\footnote{$c$ is set 1 in the paper.}, it
follows an important identity
\begin{equation}
\frac{d  E}{d { q_i^{}}}={consts}.
\end{equation}
It would mean that there is some `wave packet' moving with constant
`group velocity'. Namely, a law of inertia-like in space of momentum
hidden in Snyder's model \cite{Duality}.

Regarding such a `wave packet' as an object in the space of
momentum, a 8-d phase space $({\cal M}_a, \omega_a)$ for  can be
constructed and locally there are Snyder's momentum ${ q_\mu^{}}$
 as canonical momentum and the conjugate variables $X^\mu$ as canonical
coordinates ${(q_\mu^{}, X^{\mu}={ R}g_{a}^{\mu\nu}
{dq_\nu}/{ds_a})}$ with a symplectic  structure $\om_a$ and basic
Poisson brackets {$\{q_\mu^{}, X^\nu\}_a=-\delta^\nu_\mu$,
$\{q_\mu^{}, q_\nu^{}\}_a=0$, $\{X^\mu, X^\nu\}_a=0.$} Then Snyder's
space-time `coordinates' $x_a^\mu$ and angular momentum $l_a^{\mu
\nu}$ can be expressed in terms of these canonical variables
$(q_\mu^{}, X^{\mu})$. And it is straightforward to show that they
form an $\mathfrak{so}(1,4)$ under Poisson bracket.

In a momentum picture of the canonical quantization, the operators
of Snyder's `coordinates' and angular momentum are just ten Killing
vectors of the model, up to some coefficients,
\be\label{xt}%
\hat x_a^i :=i\hbar\left [\d {\partial}{\partial q_i^{}}-\d
{a^2}{ \hbar ^2}q^i { q_\mu^{}}\d {\partial}{\partial  q_\mu^{}}\right]
=\hat {\cal L}_a ^{4i},
 &\smallskip\\
\hat x_a^0:=i\hbar\left [\d {\partial}{\partial  q_0^{}}- \d {a^2}
{\hbar^2}q^0 {q_\mu^{}}\d {\partial}{\partial  q_\mu^{}}\right ]
 =\hat {\cal L}_a ^{40}. &
\ee%
Together with  `boost' $\hat M_{a i}=\hat x_{a}^i q_0^{}+\hat
x_a^0 q_i^{}=:\hat l_a^{0i}=\hat {\cal L}_a^{0i}$ %
and `3-angular momentum' $\hat L_{a i}= -\frac 1 2\eps_{ijk}(\hat
x_a^j q^k-\hat x_a^k q^j) =:\frac 1 2\eps_{ijk} \hat l_a^{jk} =\frac
1 2\eps_{ijk}\hat {\cal L}_a^{jk}$, they are the components of 5-d
angular momentum $\hat{\cal L}_a^{AB}$ and form an
$\mathfrak{so}(1,4)$ algebra:
\be\label{so14m}%
&&[\hat x_a^i, \hat x_a^j]=i\hbar^{
-1}{a}^{2} \hat l_a^{ij},~\quad
[\hat x_a^0, \hat x_a^i ]={i\hbar^{ -1}{a}^{2}} \hat l_a^{0i}, \\
\nno%
&&[\hat L_{a i}, \hat L_{a j}]=i\hbar\epsilon_{ij}^{\ \ k}\hat L_{a k}, \quad [\hat M_{a i}, \hat
M_{a j}]= i \hbar{\hat l_a^{ij}},\quad  \eps_{12}^{\ \ 3}=-1, ~~etc.
\ee%

Obviously, Snyder's quantized space-time `coordinates'
$\mathfrak{so}(1,4)$ algebra is the same as the coordinate
$\mathfrak{so}(1,4)$ subalgebra of Yang's $\mathfrak{so}(1,5)$. But,
the operators of canonical coordinates ${ \hat X^\mu}$ are still
commutative.

  In order to get Snyder's model
 from the complete Yang model,
 we consider a dimensionless
 $d{\cal S}_5 \cong {{\mathscr H}}\subset \mathscr{M}^{1,5}$:
\be\label{Hr/a}%
 \mathscr{H}:\quad {\eta}_{\cal{A} \cal{B}} \zeta^{\cal{A}}
\zeta^{\cal{B}} = -\frac{R^2}{a^2}.%
\ee%

Take a subspace ${\mathscr I}_1$ of $\mathscr{H}\subset
\mathscr{M}^{1,5}$ as an intersection %
\be\label{Ia}%
{\mathscr I}_1=\mathscr{H}|_{\zeta^4=0}:\quad \mathscr{H} \cap
\mathscr{P}|_{\zeta^4=0}\subset \mathscr{M}^{1,5},\ee%
where $\mathscr{P}|_{\zeta^4=0}$ is a hyperplane defined by $\zeta^4=0$.
Introduce  dimensional coordinates%
\begin{equation}
{\eta_\mu^{}=\frac{ \hbar}{R}\zeta^\mu, \qquad \eta_4^{}=\frac{
\hbar }{R}\zeta^5,}
\end{equation}
then the subspace ${\mathscr I}_1$ becomes   ${\cal H}_a$ (\ref{Ha})
with  metric (\ref{dsa}) related to the metric
(\ref{dChi}) restricted on ${\mathscr I}_1$%
 \be
ds_a^2=\frac{\hbar^2}{R^2}d\chi^2|_{{\mathscr I}_1}.  \ee%

Thus, the ${\cal A, B} \neq 4$ components of Yang's 6-d `angular
momentum' ${\mathscr L}^{\cal AB}$ consist of a 5-d angular momentum
which is identical to the 5-d angular momentum ${\cal L}_a^{AB}$ in
Snyder's model, and Yang's operators of coordinate and angular
momentum $\{\hat x^\mu,~ \hat l^{\mu \nu}\}$ and their algebra are
the same as Snyder's operators $\{\hat x_a^\mu,~ \hat
l_a^{\mu\nu}\}$ for the Killing observables and their algebra.
Therefore, the complete Yang model really contain Snyder's model as
a sub-model.

\subsection{The Beltrami model of \dS\ \SR\ from the Yang model}
 On a \dS-spacetime with radius $R$ as a hyperboloid
embedded in a 1+4-d \Mink-space %
\be\label{Hr}%
 {\cal H}_R:\quad \eta_{AB} \xi^A \xi^B = -R^2, \quad 
 ds_R^2=\eta_{AB} d\xi^A
d\xi^B, \ee%
 a free particle with mass ${ \hbar}/a$ may move uniformly along a great
`circle' defined by a conserved 5-d angular momentum
\be\label{L5R}%
 \frac{d{\cal
L}_R^{AB}}{ds_R}=0,\quad {\cal
L}_R^{AB}:=\frac {\hbar} a(\xi^A\frac{d\xi^B}{ds_R}-\xi^B\frac{d\xi^A}{ds_R}),%
\ee%
 with an Einstein-like formula for  mass ${ \hbar}/a$
\begin{eqnarray}\label{emla}%
-\frac{1}{2R^2}{\cal L}_R^{AB}{\cal L}_{R AB}=\frac { \hbar^2} {a^2},\quad%
{\cal L}_{R AB}=\eta_{AC}\eta_{BD}{\cal L}_R^{CD}.
\end{eqnarray}

 The conserved  momentum and  angular momentum of the particle can be defined as
\be%
\frac{d p_R^\mu}{ds}=0 \quad p_R^\mu=\frac{ \hbar}{Ra}\left(\xi^4
\frac{d \xi^\mu}{ds_R} - \xi^\mu
\frac{d \xi^4}{ds_R}\right), \\%
\frac{d l_R^{\mu\nu}}{ds}=0, \quad l_R^{\mu
\nu}=\frac{\hbar}{a}\left(\xi^\mu
\frac{d\xi^\nu}{ds_R}-\xi^\nu \frac{d \xi^\mu}{ds_R}\right). \ee%
And the Einstein-like formula becomes%
\begin{equation} %
p_R^\mu p_{R\mu} -\frac{1}{2R^2}l_R^{\mu\nu}l_{R\mu\nu}=\frac{
\hbar^2}{a^2},
\end{equation}
where $p_{R\mu}=\eta_{\mu\nu}p_R^\nu$ and
$l_{R\mu\nu}=\eta_{\mu\rho}\eta_{\nu\sigma}l_R^{\rho\sigma}$.

In  a Beltrami atlas of the \BdS-spacetime \cite{BdS}, the Beltrami
coordinates read in the patch $U_{4+}, \xi^4>0$, 
\begin{equation}
y^\mu=R \frac{\xi^\mu}{\xi^4},\quad \xi^4 \neq 0
\end{equation}
and the metric becomes $ds_R^2=g_{R \mu
\nu}dy^\mu dy^\nu$ with %
 \be\label{dsrB}%
 g_{R \mu
\nu} =\sigma^{-1}(y)\eta_{\mu \nu}+R^{-2} y_\mu y_\nu
\sigma^{-2}(y),\quad \sigma(y)=1-R^{-2}y^\nu y_\nu>0,
 \ee%
 where $y_\mu =\eta_{\mu\nu} y^\nu.$
For the free particle with mass $\hbar/a$ along geodesic that is the
great `circle' on ${\cal H}_R$, its motion becomes uniform motion
with constant coordinate velocity. In fact, its  momentum and
angular momentum
\be\label{prc}%
p_R^\mu={\frac{\hbar}{a}}\sigma^{-1}(y) \frac{d
y^\mu}{ds_R},\quad
 l_R^{\mu \nu}={\frac{ \hbar}{a}}(y^\mu
\frac{dy^\nu}{ds_R}-y^\nu \frac{dy^\mu}{ds_R}),
\ee
are constants. This leads to the law of inertia 
 for the particle:%
\be%
v^i:=\frac{dy^i}{dt}=consts.%
\ee%

 For the particle, there is  an associated phase space $({\cal M}_R, \omega_R)$
 and locally there are  Beltrami coordinates as the canonical
 coordinates and the covariant 4-momentum as canonical momentum
$(y^\mu,~ P_{\mu}={ \frac{\hbar}{a}}g_{\mu\nu} {dy^\nu}/{ds_R})$ with a %
symplectic structure and basic Poisson brackets in the patch 
 $\{y^\mu,
P_\nu\}_R=-\delta_\nu^\mu,~ \{y^\mu, y^\nu\}_R=0, ~
\{P_\mu, P_\nu\}_R=0.$ 
Now,  the conserved Killing momentum and angular momentum of the
particle can be expressed in terms of the canonical variables and
form an $\mathfrak{so}(1,4)$ algebra under the Poisson bracket.

In a canonical coordinate picture of the canonical quantization, the
operators of these conserved Killing Beltrami momentum and angular
momentum are just ten Killing vectors of the model up to some
coefficients forming an $\mathfrak{so}(1,4)$ under Lie bracket
\begin{equation}
\begin{split}
[\hat p_R^\mu, \hat p_R^\nu]&=\frac{i\hbar}{R^2}\hat l_R^{\mu
\nu},\quad [\hat l_R^{\mu \nu}, \hat
p_R^\rho]=i\hbar(\eta^{\nu\rho} \hat p_R^\mu -\eta^{\mu\rho} \hat
p_R^\nu), %
\end{split}
\end{equation}
together with an $\mathfrak{so}(1,3)$ for angular momentum operators
$ \hat l_R^{\mu\nu}$.  This is the same as the momentum subalgebra
of the Yang model.

It is remarkable that the conserved Killing Beltrami momentum lead
to the law of inertia in the patch and it holds globally in the
atlas patch by patch. In fact, the \dS\ \SR\ can be set up based on
the principle of inertia and the postulate of universal constants,
the speed of light $c$ and the \dS-radius $R$ \cite{BdS}.

In order to show that there is indeed the \BdS-model of the \dS\
\SR\ from the complete Yang model, let us consider another subspace
${\mathscr I}_2$ of $\mathscr{H}\subset
\mathscr{M}^{1,5}$ (\ref{Hr/a})  of the Yang model as an intersection%
\be\label{Ir}%
{\mathscr I}_2=\mathscr{H}|_{\zeta^5=0}:\quad \mathscr{H} \cap
\mathscr{P}|_{\zeta^5=0}\subset \mathscr{M}^{1,5},\ee%
where $\mathscr{P}|_{\zeta^5=0}$ is a hyperplane defined by $\zeta^5=0$.  Introduce dimensional coordinates%
\begin{equation}%
\xi^\mu= a \zeta^\mu \qquad \xi^4=a \zeta^4,
\end{equation}
then ${\mathscr I}_2$ becomes the \dS-hyperploid ${\cal H}_R$
(\ref{Hr}) and its metric (\ref{dsrB}) becomes the metric
(\ref{dChi}) restricted on
${\mathscr I}_2$%
 \be\label{dChi2}%
ds_R^2=a^2 d\chi^2|_{{\mathscr I}_2}.\quad \ee%

It is also straightforward now to find that the ${\cal A, B}\neq 5$
components of the  6-d `angular momentum' operators $\hat {\mathscr
L}^{\cal AB}$ in the Yang model consist of a 5-d angular momentum,
which is just the angular momentum
operators $\hat {\cal L}_R^{AB}$ in the \dS\ special relativity. 
And Yang's momentum, angular momentum operators $\hat p^\mu, \hat
l^{\mu \nu}$ in (\ref{yang operators}) and their subalgebra are just
the Killing Beltrami momentum, angular momentum operators $\hat
p_R^\mu, \hat l_R^{\mu\nu}$ in the Beltrami model of the \dS\ \SR.
Thus,  the complete Yang model really contain the \BdS-model of \dS\
\SR\ as a sub-model.

\subsection{The Snyder's model--\dS\ \SR\ duality in the Yang model}

 It is important to see \cite{Duality} that   between Snyder's model and the
  \dS\ \SR, there is also a $\mathbb{Z}_2=\{e,s| s^2=e\}$
  dual exchange with
\be\label{dual}%
{s}:\quad x_a^\mu \rightarrow p_R^\mu,
\quad a \rightarrow \frac{\hbar}{R}. \ee%
 This is also a UV-IR
exchange. And it is isomorphic to the $Z_2$ duality in Yang's
model (\ref{duality}).

The Snyder's model-\dS\ \SR\ duality contains some other contents.
One is that the cosmological constant $\Lambda$ should be a
fundamental constant in the Nature like $c, G$ and $\hbar$. This
is already indicated in Yang's model as long as
$R=(3/\Lambda)^{1/2}$ is taken.

 Thus, not only  both Snyder's model and the \dS\ \SR\ are
sub-models in the complete Yang model, but their $\mathbb{Z}_2$
duality transformations are contained in that of  the $Z_2$ duality
in the Yang model as well.

\section{Concluding Remarks}

The above `surgery' for two subspaces ${\mathscr I}_2$ and
${\mathscr I}_2$  of a dimensionless $\mathscr{H}\subset
\mathscr{M}^{1,5}$ (\ref{Hr/a})  in the Yang model shows that
  both Snyder's model and  the \BdS-model of  \dS\ \SR\ are really
  sub-physics of the complete Yang model. And
the UV-IR duality in the Yang model is just the one-to-one exchange
of the Snyder model--\dS\ \SR\ duality.

It is quite possible that there are some other physical
implications and/or relations with other dualities, such as the
T-duality and S-duality, if $a$ and $R$ may have other
identifications.

 It should be mentioned that a Yang-like model with an
$\mathfrak{so}(2,4)$ symmetry on a dimensionless $(2,4)$-d flat
space $\mathscr{M}^{2,4}$ can be
 set up and all similar
issues for Snyder's model and the \dS\ \SR\ or for an anti-Snyder's
model on an \AdS-space of momentum and the \AdS\ \SR\ can also be
realized started with a dimensionless \AdS$_5 \cong \mathscr{H}$ or
its boundary $\partial ({A}d{S}_5) \cong \mathscr{N}\subset
\mathscr{M}^{2,4}$.

Since Yang's algebra is just the Lie algebra form of the deformed
algebra in TSR, which is a generalization of DSR, it should be
explored from the point of view in our approach what are the
relations with the other DSR models and the TSR.  It seems that  the
duality exists between other DSR models and \dS-spacetime since some
DSR models can be realized as Snyder's model in different
coordinates on the \dS-space of momentum, the corresponding
coordinates on the \dS-spacetime may also be taken. But, this may
cause some issues due to the non-inertial effects from the viewpoint
of the \dS\ \SR.

Finally, we would like to emphasize that the complete Yang model
should be regarded as a theory of the \SR\  based on the principle
of inertia in the both spacetime and space of momentum as well as
the postulate on three universal constants $c, \ell_P$ and $R$.

All above  issues and related topics should be further
studied.

\begin{acknowledgments}
This work is completed during the string theory and cosmology program, KITPC,
CAS.  We would like to thank Z. Chang, Y. Ling, Q. K. Lu, X. C. Song, Y. Tian,
S. K. Wang, K. Wu, X. N. Wu, Y. Yang, Z. Xu and B. Zhou
 for valuable discussions. This work is partly supported
by NNSFC (under Grant Nos. 90403023, 90503002, 10701081, 10775140), NKBRPC
(2004CB318000), Beijing Jiao-Wei Key project (KZ200810028013) and Knowledge
Innovation Funds of CAS (KJCX3-SYW-S03).
\end{acknowledgments}

\end{document}